\documentclass[12pt]{article}
\usepackage[symbol]{footmisc}
\def\correspondingauthor{\footnote{Corresponding author.  }}
\usepackage[left=2.5cm,top=2.5cm,right=2cm,bottom=2.5cm]{geometry}
\usepackage{amsmath, amssymb}

\usepackage{graphicx}
\usepackage{graphics}
\usepackage{epstopdf}
\usepackage{subfigure}
\usepackage{amsfonts}
\usepackage{sectsty}
\usepackage{sectsty}
\usepackage{hyperref}
\usepackage{cite}

\begin{document}
	\begin{center}
	\large{\bf{Static Traversable Wormholes in $f(R, T)=R+2\alpha \ln T$ Gravity}} \\
	\vspace{5mm}
\normalsize{Nisha Godani$^1$ and Gauranga C. Samanta$^{2}{}$\correspondingauthor{} }\\
\normalsize{$^1$Department of Mathematics, Institute of Applied Sciences and Humanities\\ GLA University, Mathura, Uttar Pradesh, India\\
$^2$Department of Mathematics, BITS Pilani K K Birla Goa Campus, Goa, India}\\
\normalsize {nishagodani.dei@gmail.com\\gauranga81@gmail.com}
\end{center}
	
	\begin{abstract}
Traversable wormholes, studied by Morris and Thorne \cite{Morris1} in general relativity,  are investigated in this research paper in $f(R,T)$ gravity by introducing a new form of non-linear $f(R,T)$ function. By using this novel function, the Einstein's field equations in $f(R,T)$ gravity are derived. To obtain the exact wormhole solutions, the relations $p_t=\omega\rho$ and $p_r=\sinh(r)p_t$, where $\rho$ is the energy density, $p_r$ is the radial pressure and $p_t$ is the tangential pressure, are used. Other than these relations, two forms of shape function defined in literature are used, and their suitability is examined by exploring the regions of validity of null, weak, strong and dominant energy conditions . Consequently, the radius of the throat or the spherical region, with satisfied energy conditions, is determined and the presence of exotic matter is minimized.
\end{abstract}

\section{Introduction}
Recently, cosmologists are trying to construct an exact traversable wormhole model without exotic matter. However, it is a challenging problem not only in general relativity, but also in modified theories. Now, the question lies,  whether the laws of physics are allowed to construct and sustain wormholes for interstellar travel. Such a wormhole is a tunnel in the topology of space, which links widely two separated regions of the universe. This type of geometry could be described by the Schwarzschild metric with an appropriate choice of topology\cite{Wheeler1, Misner1973}. However,
the Schwarzschild wormhole contains the horizon. It avoids two-way travel. Its throat squeezes so quickly that
it cannot be traversed in even one direction\cite{Misner1973}. To avoid horizons and singularities, one must thread the throat with nonzero stress and energy \cite{Morris1}. Now, the researchers could face two questions: (a) Does quantum field theory allow such kind of stress-energy tensor which is essential to sustain a two-way traversable wormhole? (b) Do the laws of physics allow the construction of wormholes in a universe as its spatial sections initially are simply connected? These questions become important when one identifies that the laws of physics allow traversable wormhole. They perhaps allow such a wormhole to be transformed into a "time machine" with which causality might be violated.

The wormholes are tunnel like structures that possess non-trivial topology and connect two space-times or two different points in the same space-time. Einstein and Rosen \cite{eins-ros} obtained the wormhole geometry known as Einstein-Rosen bridge that connects two sheets. Wheeler \cite{Wheeler} obtained Reissner-Nordstr\"{o}m or Kerr wormholes in the form of geometries of quantum
foam at the Planck scale. Hawking   \cite{Hawking} transformed these wormholes into Euclidean wormholes through which, the travel from one side to the other side was not possible. Later, Thorne with his student Morris  \cite{Morris1} constructed traversable wormholes with two mouths and one throat. They considered static and spherically symmetric wormholes, performed the meticulous study and obtained the presence of the exotic matter that violates the energy conditions. They presented a class of wormhole solutions of the Einstein field equation  which described that human beings could be traversed through the tunnel of the wormholes, provided, there should not be any horizon throughout the tunnel. So, near the throat of the wormhole the material must hold the radial tension greater than the mass energy density $(\tau_0>\rho_0c^2)$. So far, no known material has this $\tau_0>\rho_0c^2$ property and such type of material would violates all energy conditions. For backward time travel, these wormholes can also be converted into time machines \cite{Morris2, Frolov}.

The traversable wormholes do not satisfy all energy conditions simultaneously in general theory of relativity\cite{Morris1}. However, there are some gravitational theories in which a traversable wormhole could follow the energy conditions\cite{Mehdizadeh04, Zangeneh49, Mehdizadeh14, Shaikh11, Moradpour66}. In fact, investigation of various traversable wormholes in different modified theories is a significant and prominent issue in theoretical physics. The general theory of relativity could be modified in many different aspects.  The result of these modifications reveals the large number of modified theories available in the literature \cite{Tsujikawa, Capozziello,  Nojiri, Clifton, Berti}. The $f(R)$ modified gravity \cite{Buchdahl} attains ample consideration for its capability to elucidate the expansion of the universe. In the early 1980s, Starobinsky \cite{Star} discussed $f(R)$ model by taking $f(R)=R+\alpha R^2$, where $\alpha>0$, represents inflationary scenario of the universe.
The $f(R)$ theory of gravity replaces the scalar curvature $R$ in the Einstein gravitational
action by an arbitrary function $f(R)$.
A simplification of $f(R)$ gravity suggested in \cite{Bertolami} integrates an unambiguous coupling between the matter Lagrangian and an arbitrary function of the scalar curvature, which leads to an extra force in the geodesic equation of a perfect fluid. Subsequently, it is shown that this extra force may be a justification for the accelerated expansion of the universe \cite{Nojiris, Bertolami1, Bamba}.
The dynamical behavior of the matter and the dark energy effects have been obtained within extended theories of gravity \cite{Nojiri001, Shirasaki, Capozziello001,  Rodrigues01}. Apart from this, recently  many scientists have studied the dynamics of cosmological models using $f(R)$ gravity in various directions\cite{Capozziello91, Bombacigno21, Sbis, Chen, Elizalde26, Elizalde06, Astashenok1, Miranda, Nascimento, Odintsov20, Odintsov1, Nojiri56, Parth, gninjp}.  Unsurprisingly, these models may also have some inadequacies. For example, solar system tests have ruled out most of the $f(R)$ models suggested so far \cite{Chiba, Nojiri21}.

Apart from this, $f(R, T)$ theory \cite{harko} has also gained much attention to explain accelerated expansion of the universe.
In this theory the matter term is included in gravitational action, in which the gravitational Lagrangian
density is an arbitrary function of both $R$ and $T$, where $R$ is the Ricci scalar and $T$ is the trace of the stress-energy tensor.
The random requirement on $T$ summarizes the conceivable contributions from both non-minimal coupling and unambiguous $T$ terms.
Many functional forms of $f(R, T)$ theories have been studied for the effect of cosmological dynamics in different contexts.
The split-up $f(R, T)=f_1(R)+f_2(T)$, where $f_1$ and $f_2$ are arbitrary functions of $R$  and $T$ respectively, has established much attention because one can explore the contributions from
$R$ without specifying $f_2(T)$.  Similarly one can explore the contributions from $T$ without specifying $f_1(R)$.
In such separable theories, the reconstruction of $f(R, T)$ gravity is studied\cite{Houndjo1}.
A non-equilibrium picture of thermodynamics at the apparent horizon of the (FLRW) universe is studied\cite{Sharif01}. Several authors have adopted the form $f(R, T)=f_1(R)+f_2(T)$ to  study the cosmological dynamics from different aspects\cite{Jamil1, Alvarenga01, Santos1, Samanta, Shabani, Naidu, Samanta3, Chandel, Samanta1, Shabani001, Samanta2, Moraes01, Noureen, Farasat, Mirza, Correa1, Ramesh, Moraes0001, Moraes9, Zaregonbadi, Rahaman1, Mishra, Yousaf, Samanta5, Aditya, Elizalde, Ordines}.

Many cosmologists have studied wormhole solutions in modified theories of gravity from different aspects. We are going to discuss the contributions of some cosmologists in $f(R)$ and $f(R,T)$ gravity theories.
Pavlovic and Sossich \cite{Pavlovic} introduced a  variable redshift function and obtained wormhole solutions without exotic matter in $f(R)$ gravity.
Mazharimousavi and  Halilsoy \cite{Habib92} obtained wormhole solutions satisfying weak energy condition in $f(R)$ gravity.
Lobo and Oliveira \cite{Lobo1} obtained wormhole solutions in $f(R)$ gravity using specific shape functions and constant redshift function and explored the factors supporting the wormhole structures.
 Saiedi and Esfahani \cite{Saiedi} used constant redshift and  shape functions, and examined 'null and weak' energy conditions to  obtain the wormhole solutions in $f(R)$ gravity.
 Eiroa and Aguirre \cite{Eiroa1} studied thin-shell wormholes with a charge in $f(R)$ gravity and checked their stability under perturbations.
Kuhfittig \cite{peter}, using the framework of $f(R)$ gravity, derived $f(R)$ functions for different shape functions. They also obtained wormhole solutions for specific choices of $f(R)$ functions. Godani and Samanta \cite{ns} studied traversable wormhole in $f(R)$ gravity for two specific forms of shape functions and found wormhole structures are filled with phantom fluid. Samanta et al. \cite{godani1} defined an exponential shape function and compared the validity of energy conditions in $f(R)$ and $f(R,T)$ theories of gravity. Recently, Godani and Samanta \cite{ngmpla} and Samanta and Godani \cite{gnmpla} found the satisfaction of energy conditions for a wide range of radial coordinate in the context of traversable wormholes in $f(R)$ gravity. Further,  Samanta and Godani \cite{gnepjc} having considered power law shape function and an equation of state, derived a viable $f(R)$ model and obtained the validity of energy conditions.

Azizi \cite{Azizi} derived shape function and obtained the wormhole solutions satisfying the null energy condition using the background of $f(R, T)$ gravity.
Zubair et al. \cite{Zubair} considered static and spherically symmetric wormholes with three types of fluids with $f(R, T)=f(R)+\lambda T$ gravity, where $\lambda$ is a constant one. They used  Starobinsky $f(R)$ model and analyzed the energy conditions.
Zubair et al. \cite{Zubair80} proposed non-commutative geometry in terms of Gaussian and Lorentizian distributions of string theory and obtained the exact and numerical solutions in $f(R,T)$ gravity. They also discussed gravitational lensing for exact wormhole solutions.
Yousaf et al. \cite{Yousaf63} considered wormholes filled with two different fluids and obtained solutions without the presence of exotic matter.
Moraes et al. \cite{Moraes2} used an analytical approach to obtain the wormhole solutions in $f(R, T)$ gravity.
Elizalde and Khurshudyan \cite{Elizalde25} studied wormholes in $f(R,T)$ gravity for two forms of varying Chaplygin gas and obtained the violation of null and dominated energy conditions.
Bhatti et al. \cite{Bhatti} considered exponential $f(R,T)$ gravity model and found wormhole solutions.
E. Elizalde and M. Khurshudyan \cite{Elizalde51} investigated traversable wormhole solutions  in $f(R,T)$ gravity by considering the various forms of energy density.
Moraes et al. \cite{Moraes1} studied the charged wormhole solutions in $f(R, T)$ theory of gravity and found satisfaction of energy conditions.
Sharif and  Nawazish \cite{Sharif} investigated wormhole solutions for dust and non-dust distributions using  Noether symmetry approach in $f(R, T)$ gravity. They considered both the variable and constant forms of redshift function and found the existence of stable and traversable wormhole solutions.

The motivation of this paper is to study the traversable wormholes in $f(R, T)$ gravity by defining new form of $f(R, T)=R+2\alpha \ln T$
function to minimize the presence of exotic type of matter near the throat of the wormhole. Therefore, in this work,  we have assumed logarithmic form for the dependence on $T$ in $f(R,T)$ gravity. The regions satisfying the energy conditions are explored. Consequently, the wormhole solutions without exotic matter are obtained.

\section{The Background of $f(R, T)$ Gravity}
The $f(R, T)$ theory of gravity was first introduced by Harko et al. \cite{harko} in $2011$. They extended standard general theory of relativity by modifying gravitational Lagrangian. The gravitational action in $f(R, T)$ theory is given by
\begin{equation}\label{action2}
S=S_G + S_m=\dfrac{1}{16\pi}\int f(R,T)\sqrt{-g}d^4x +\int \sqrt{-g}\mathcal{L} d^4x,
\end{equation}
where $f(R, T)$ is assumed to be an arbitrary function of $R$ and $T$. Precisely, $R$ is the Ricci scalar and $T$ is the trace of the energy momentum tensor $T_{\mu\nu}$. The matter Lagrangian density is denoted by $\mathcal{L}$ and the energy momentum tensor is defined in terms of the matter action as follows \cite{Landau}:

\begin{equation}\label{}
  T_{\mu\nu}=-\frac{2\delta (\sqrt{-g}\mathcal{L})}{\sqrt{-g}\delta g^{\mu\nu}},
\end{equation}
which yields
\begin{equation}\label{}
  T_{\mu\nu}=g_{\mu\nu}\mathcal{L}-2\frac{\partial \mathcal{L}}{\partial g^{\mu\nu}}.
\end{equation}
The trace $T$ is defined as $T=g^{\mu\nu}T_{\mu\nu}$. Let us define the variation of $T$ with respect to the metric
tensor as
\begin{equation}\label{}
  \frac{\delta (g^{\alpha\beta}T_{\alpha\beta})}{\delta g^{\mu\nu}}=T_{\mu\nu}+\Theta_{\mu\nu},
\end{equation}
where $\Theta_{\mu\nu}=g^{\alpha\beta}\frac{\delta T_{\alpha\beta}}{\delta g^{\mu\nu}}$.
Varying action (\ref{action2}) with respect to the metric tensor $g^{\mu\nu}$, yields
\begin{equation}\label{frt}
f_R(R,T)R_{\mu\nu} -\frac{1}{2}f(R,T)g_{\mu\nu} + (g_{\mu\nu}
\square
-\triangledown_\mu\triangledown_\nu)f_R(R,T)=8\pi T_{\mu\nu} -f_T(R, T)T_{\mu\nu}- f_T(R,T) \Theta_{\mu\nu},
\end{equation}
where $f_R(R,T) \equiv \dfrac{\partial f(R, T)}{\partial R}$ and
$ f_T(R,T) \equiv \dfrac{\partial f(R,T)}{\partial T}.$ Note that, if we take $f(R, T)=R$ and $f(R, T)=f(R)$, then the equations \eqref{frt} becomes Einstein field equations of general relativity and $f(R)$ gravity respectively.

 In this present study, we consider $f(R, T)=R+2f(T)$, where $f(T)$ is an arbitrary function of trace of the energy momentum tensor $T$. Now, using this particular form of $f(R, T)$ function in equations \eqref{frt}, we obtained
\begin{equation}\label{field}
  R_{\mu\nu}-\frac{1}{2}R g_{\mu\nu}=8\pi T_{\mu\nu}-2f'(T)T_{\mu\nu}-2f'(T)\Theta_{\mu\nu}+f(T)g_{\mu\nu},
\end{equation}
where prime stands for derivative with respect to the argument. If the matter source is a perfect fluid, $\Theta_{\mu\nu}=-2T_{\mu\nu}-pg_{\mu\nu}$, then the field equations  \eqref{field} become
\begin{equation}\label{ffield}
  R_{\mu\nu}-\frac{1}{2}R g_{\mu\nu}=8\pi T_{\mu\nu}+2f'(T)T_{\mu\nu}+(2pf'(T)+f(T))g_{\mu\nu}.
\end{equation}
Note that, if we take $f(T)\equiv 0$, then the field equations \eqref{ffield} become general Einstein field equations.

\section{$f(R, T)=R+2\alpha\ln(T)$ and Wormholes Modeling}
In the present paper, we define a novel form of $f(R,T)$ function as
\begin{equation}\label{newfunc}
  f(R, T)=R+2\alpha\ln(T),
\end{equation}
where $\alpha$ is constant and $T=-\rho+p_r+2p_t$. From the above choice of $f(R, T)$, the term $T=-\rho+p_r+2p_t$ should be greater than zero,  otherwise the function $f(R, T)$ will not be well defined. Therefore, the condition $T=-\rho+p_r+2p_t>0$ is mandatory.
In this functional form, the logarithmic dependence on $T$ is new in literature. Models exponentially depending on $R$ or $T$ terms are defined in literature in the context of both $f(R)$ and $f(R,T)$ gravity theories \cite{Linder, Campista, Bamba1, Odintsov}. Since logarithmic and exponential functions are inverse of each other and, wormhole solutions have been obtained for exponential dependence on $T$ in $f(R,T)$ function, therefore a natural question arises: whether the wormhole solutions with or without exotic matter exist for logarithmic dependence on $T$ in $f(R,T)$. The main motivation is to minimize the presence of exotic matter near the throat of the wormhole. Hence, this is the reason for defining $f(R,T)$ function in terms of logarithmic function of $T$. \\

A static and spherically symmetric wormhole structure is defined by the metric
\begin{equation}\label{metric}
ds^2=-e^{2\Phi(r)}dt^2+\frac{dr^2}{1-b(r)/r} + r^2(d\theta^2+\sin^2\theta d\phi^2),
\end{equation}
where the functions $b(r)$ and $e^{2\Phi(r)}$ are called as shape and redshift functions respectively.
The radial coordinate $r$ varies from $r_0\ne 0$ to $\infty$, where $r_0$ is called the radius of the throat. The angles $\theta$ and $\phi$ vary from 0 to $\pi$ and 0 to $2\pi$ respectively. To avoid the presence of horizons and singularities, the redshift function should be finite and non-zero.  The shape function should satisfy the following properties: (i) $\frac{b(r)}{r}<1$ for $r>r_0$,  (ii) $b(r_0)=r_0$ at $r=r_0$, (iii) $\frac{b(r)}{r}\rightarrow 0$ as $r\rightarrow \infty$, (iv) $\frac{b(r)-b'(r)r}{b(r)^2}>0$ for $r>r_0$ and (v) $b'(r_0)\leq1$.

The energy momentum tensor for the matter source of the wormholes is $T_{\mu\nu}=\frac{\partial L_m}{\partial g^{\mu\nu}}$, which is defined as
\begin{equation}
T_{\mu\nu} = (\rho + p_t)u_\mu u_\nu - p_tg_{\mu\nu}+(p_r-p_t)X_\mu X_\nu
\end{equation}	
such that
\begin{equation}
u^{\mu}u_\mu=-1 \mbox{ and } X^{\mu}X_\mu=1,
\end{equation}

where $\rho$,  $p_t$ and $p_r$  stand for the energy density, tangential pressure and radial pressure respectively.

The  Ricci scalar $R$ is given by $R=\frac{2b'(r)}{r^2}$ and the field equations for the metric \eqref{metric} in  $f(R, T)=R+2\alpha\ln(T)$ gravity are obtained as:

\begin{eqnarray}
\frac{b'(r)}{r^2}&=&8\pi\rho-2\alpha-\alpha\ln(-\rho+p_r+2p_t)\label{f1a}\\
\frac{-b(r)}{r^3}&=&8\pi p_r +\frac{4\alpha(p_r+p_t)}{-\rho+p_r+2p_t} +\alpha\ln({-\rho+p_r+2p_t})\label{f1b}\\
\frac{b(r)-rb'(r)}{2r^3}&=&8\pi p_t +\frac{2\alpha(3p_t+p_r)}{-\rho+p_r+2p_t} +\alpha\ln({-\rho+p_r+2p_t})\label{f1c},
\end{eqnarray}
where the prime stands for the differentiation with respect to the radial co-ordinate $r$.

\section{Wormholes Solutions}
In Section 3, the novel $f(R,T)$ function is defined as $f(R,T)=R+2\alpha\ln(T)$, where $\alpha$ is constant and $T=-\rho+p_r+2p_t>0$. Using this $f(R,T)$ function, the field equations \eqref{f1a}-\eqref{f1c} are derived. We have chosen function $b(r)$ and relation between  $\rho$, $p_r$ and $p_t$ to find out the exact solutions. In literature, several cosmologists have considered the equation of state $p_r=\omega \rho$. In our study, the field equations are highly non-linear in terms of  energy density and pressures. The choice $p_r=\omega \rho$ makes the field equations and hence the solutions very complicated. Therefore, in the present work, we have considered the equation of state
\begin{equation}\label{eos}
p_t=\omega \rho,
\end{equation}
where $\omega$ is the equation of state parameter.
Since both radial and tangential pressures may be isotropic or anisotropic, we have assumed the relation between these two pressures as
\begin{equation}\label{pre}
p_r=kp_t,
\end{equation}
where $k$ can be constant or variable.

Using equations \eqref{eos} and \eqref{pre} in\eqref{f1a}-\eqref{f1c} and performing some algebra,  the equation of state parameter $\omega$ comes out to be equal to
\begin{equation}\label{omega}
\omega = \frac{r b'(r)-3 b(r)}{(k-2) r b'(r)+(k+2) b(r)}.
\end{equation}
Further, using this value of $\omega$, the tangential pressure $p_t$ comes out to be equal to
\begin{equation}\label{pt}
p_t= \frac{\left(r b'(r)-3 b(r)\right) \left(-r b'(r)+(k+2) b(r)+\alpha  (k-1) r^3\right)}{16 \pi  (k-1) r^3 \left((k+2) b(r)-r b'(r)\right)}.
\end{equation}
If we consider isotropic pressures, i.e. $k=1$, then
$p_t$ is equal to infinity and hence $p_r$ is also equal to infinity. This implies that the value of $k$ cannot be equal to 1 or the radial and tangential pressures cannot be equal for our model. Hence, the isotropic nature of geometry could be rolled out for $f(R, T)=R+\alpha\ln T$ gravity. Now, there can be many choices for $k$ other than 1. For our model, we consider variable $k$ defined as
\begin{equation}\label{k}
k=\sinh(r).
\end{equation}

Using the value of $k$ from Eq. \eqref{k} in Equations \eqref{omega} and \eqref{pt},

\begin{equation}
\omega=\frac{r b'(r)-3 b(r)}{r (\sinh (r)-2) b'(r)+b(r) (\sinh (r)+2)}
\end{equation}
 \begin{equation}
p_t= \frac{\left(r b'(r)-3 b(r)\right) \left(-r b'(r)+b(r) (\sinh (r)+2)+\alpha  r^3 (\sinh (r)-1)\right)}{16 \pi  r^3 (\sinh (r)-1) \left(b(r) (\sinh (r)+2)-r b'(r)\right)}.
 \end{equation}
Now, to determine the exact solution, it is required to define the shape function $b(r)$ explicitly. In this work, we have considered two forms of $b(r)$ defined in the literature. One form is

   \begin{equation}
 b(r)=\frac{r_0\log(r+1)}{\log(r_0+1))},
  \end{equation}\\
proposed by Godani and Samanta \cite{ns} to analyse the wormhole solutions in $f(R)$ theory of gravity.

Another form is
  \begin{equation}
  b(r)=\frac{r}{\exp(r-r_0)},
  \end{equation}\\
 introduced by Samanta et al.\cite{godani1} for the investigation of the wormhole solutions in the background of modified $f(R)$ and $f(R,T)$ gravity theories.
  \\

Using these two forms of $b(r)$, we have obtained wormhole solutions and computed different combinations of energy density, radial and tangential pressures  in the following two cases:\\

\noindent
Case I: $ b(r)=\frac{r_0\log(r+1)}{\log(r_0+1))}$
\\
\begin{eqnarray}
\omega&=&\frac{r-3 (r+1) \log (r+1)}{-2 r+2 (r+1) \log (r+1)+(r+(r+1) \log (r+1)) \sinh (r)}
\end{eqnarray}

\begin{eqnarray}\label{pt1}
p_t&=&\frac{1}{16 \pi  (\sinh (r)-1)}
\Bigg[-\frac{3 {r_0} \log (r+1)}{r^3 \log ({r_0}+1)}+\frac{{r_0}}{r^2 (r+1) \log ({r_0}+1)}\nonumber\\
&-&\frac{\alpha  (3 (r+1) \log (r+1)-r) (\sinh (r)-1)}{-r+2 (r+1) \log (r+1)+(r+1) \log (r+1) \sinh (r)}\Bigg]\end{eqnarray}

\begin{eqnarray}\label{pr1}
p_r&=& \frac{\sinh(r)}{16 \pi  (\sinh (r)-1)}
\Bigg[-\frac{3 {r_0} \log (r+1)}{r^3 \log ({r_0}+1)}+\frac{{r_0}}{r^2 (r+1) \log ({r_0}+1)}\nonumber\\
&-&\frac{\alpha  (3 (r+1) \log (r+1)-r) (\sinh (r)-1)}{-r+2 (r+1) \log (r+1)+(r+1) \log (r+1) \sinh (r)}\Bigg]
\end{eqnarray}

\begin{eqnarray}\label{density1}
\rho&=&\frac{1}{16 \pi  r^3 (r+1) (\sinh (r)-1) \log ({r_0}+1) (-r+2 (r+1) \log (r+1)+(r+1) \log (r+1) \sinh (r))}\nonumber\\
&\times&\Bigg[(-2 r+2 (r+1) \log (r+1)+(r+(r+1) \log (r+1)) \sinh (r)) \left((r+1) \left(2 {r_0} \log (r+1)\right.\right.\nonumber\\
&-&\left.\left.\alpha  r^3 \log ({r_0}+1)\right)+(r+1) \sinh (r) \left(\alpha  r^3 \log ({r_0}+1)+{r_0} \log (r+1)\right)-r {r_0}\right)\Bigg]
\end{eqnarray}

Using Equations \eqref{pt1}, \eqref{pr1} and \eqref{density1}, we obtained the following terms:

\begin{eqnarray}
\rho+p_r&=& -\frac{1}{8 \pi  r^3 (r+1) \log ({r_0}+1) (-r+2 (r+1) \log (r+1)+(r+1) \log (r+1) \sinh (r))}\nonumber\\
&\times&\Bigg[((r+1) \log (r+1)-r) \left((r+1) \sinh (r) \left(\alpha  r^3 \log ({r_0}+1)+{r_0} \log (r+1)\right)\right.\nonumber\\
&-&r \left.\left(\alpha  (r+1) r^2 \log ({r_0}+1)+{r_0}\right)+2 (r+1) {r_0} \log (r+1)\right)\Bigg]
\end{eqnarray}

\begin{eqnarray}
\rho+p_t&=&\frac{1}{16 \pi  r^3 (r+1) \log ({r_0}+1) (-r+2 (r+1) \log (r+1)+(r+1) \log (r+1) \sinh (r))}\nonumber\\
&\times&\Bigg[(r+(r+1) \log (r+1)) \left((r+1) \sinh (r) \left(\alpha  r^3 \log ({r_0}+1)+{r_0} \log (r+1)\right)\right.\nonumber\\
&-&\left.r \left(\alpha  (r+1) r^2 \log ({r_0}+1)+{r_0}\right)+2 (r+1) {r_0} \log (r+1)\right)\Bigg]
\end{eqnarray}

\begin{eqnarray}
\rho+p_r+2p_t&=&\frac{-1}{8 \pi  r^3 (r+1) (\sinh (r)-1) \log ({r_0}+1) (-r+2 (r+1) \log (r+1)+(r+1) )}\nonumber\\
&\times&\frac{1}{\log (r+1)\sinh (r)}\Bigg[(2 (r+1) \log (r+1)+((r+1) \log (r+1)-r) \sinh (r)) \nonumber\\
&\times& \left((r+1)
\sinh (r) \left(\alpha  r^3 \log ({r_0}+1)+{r_0} \log (r+1)\right)-r \left(\alpha  (r+1) r^2 \log ({r_0}+1)\right.\right.\nonumber\\
&+&\left.\left.{r_0}\right)+2 (r+1) {r_0} \log (r+1)\right)\Bigg]
\end{eqnarray}

\begin{eqnarray}
\rho-|p_r|&=& \frac{1}{16 \pi  r^3 (r+1) (\sinh (r)-1) \log ({r_0}+1) (-r+2 (r+1) \log (r+1)+(r+1) \log (r+1) \sinh (r))}\nonumber\\
&\times&\Bigg[(-2 r+2 (r+1) \log (r+1)+(r+(r+1) \log (r+1)) \sinh (r)) \left((r+1) \left(2 {r_0} \log (r+1)\right.\right.\nonumber\\
&-&\left.\left.\alpha  r^3 \log ({r_0}+1)\right)+(r+1) \sinh (r) \left(\alpha  r^3 \log ({r_0}+1)+{r_0} \log (r+1)\right)-r {r_0}\right)\Bigg]\nonumber\\
&-&\Bigg| \frac{\sinh(r)}{16 \pi  (\sinh (r)-1)}
\Bigg[-\frac{3 {r_0} \log (r+1)}{r^3 \log ({r_0}+1)}+\frac{{r_0}}{r^2 (r+1) \log ({r_0}+1)}\nonumber\\
&-&\frac{\alpha  (3 (r+1) \log (r+1)-r) (\sinh (r)-1)}{-r+2 (r+1) \log (r+1)+(r+1) \log (r+1) \sinh (r)}\Bigg]\Bigg|
\end{eqnarray}

\begin{eqnarray}
\rho-|p_t|&=& \frac{1}{16 \pi  r^3 (r+1) (\sinh (r)-1) \log ({r_0}+1) (-r+2 (r+1) \log (r+1)+(r+1) \log (r+1) )}\nonumber\\
&\times&\frac{1}{\sinh (r)}\Bigg[(-2 r+2 (r+1) \log (r+1)+(r+(r+1) \log (r+1)) \sinh (r)) \left((r+1) \right.\nonumber\\
&\times&\left.\left(2 {r_0} \log (r+1)-\alpha  r^3 \log ({r_0}+1)\right)+(r+1) \sinh (r) \left(\alpha  r^3 \log ({r_0}+1)+{r_0} \log (r+1)\right)\right.\nonumber\\
&-&\left.r {r_0}\right)\Bigg]-\Bigg|\frac{1}{16 \pi  (\sinh (r)-1)}
\Bigg[-\frac{3 {r_0} \log (r+1)}{r^3 \log ({r_0}+1)}+\frac{{r_0}}{r^2 (r+1) \log ({r_0}+1)}\nonumber\\
&-&\frac{\alpha  (3 (r+1) \log (r+1)-r) (\sinh (r)-1)}{-r+2 (r+1) \log (r+1)+(r+1) \log (r+1) \sinh (r)}\Bigg]\Bigg|
\end{eqnarray}

\begin{eqnarray}
p_t-p_r&=&\frac{1}{16 \pi  r^3 (r+1) \log ({r_0}+1) (-r+2 (r+1) \log (r+1)+(r+1) \log (r+1) \sinh (r))}\nonumber\\
&\times&\Bigg[(3 (r+1) \log (r+1)-r) \left((r+1) \sinh (r) \left(\alpha  r^3 \log ({r_0}+1)+{r_0} \log (r+1)\right)\right.\nonumber\\
&-&r \left.\left(\alpha  (r+1) r^2 \log ({r_0}+1)+{r_0}\right)+2 (r+1) {r_0} \log (r+1)\right)\Bigg]
\end{eqnarray}

\begin{eqnarray}
-\rho+p_r+2p_t&=&\frac{1}{4 \pi  r^3 (r+1) (\sinh (r)-1) \log (r_0+1)}\Bigg[\alpha  (r+1) r^3 \log (r_0+1)-(r+1) \nonumber\\
&\times& \sinh (r)\left(\alpha  r^3 \log (r_0+1)+r_0 \log (r+1)\right)+r r_0-2 (r+1) r_0 \log (r+1)\Bigg]\nonumber\\
\end{eqnarray}

Case II: $b(r)=\frac{r}{\exp(r-r_0)}$\\
 \begin{eqnarray}
\omega&=& \frac{r+2}{(r-2) \sinh (r)-2 r}
 \end{eqnarray}

 \begin{eqnarray}\label{pt2}
 p_t&=&\frac{1}{16 \pi }\Bigg[(r+2) \left(-\frac{e^{{r_0}-r}}{r^2 (\sinh (r)-1)}-\frac{\alpha }{r+\sinh (r)+1}\right)\Bigg]
 \end{eqnarray}

 \begin{eqnarray}\label{pr2}
 p_r&=&\frac{(r+2) \sinh (r)}{16 \pi } \left(-\frac{e^{{r_0}-r}}{r^2 (\sinh (r)-1)}-\frac{\alpha }{r+\sinh (r)+1}\right)
 \end{eqnarray}

 \begin{eqnarray}\label{density2}
 \rho&=&\frac{((r-2) \sinh (r)-2 r)}{16 \pi } \left(-\frac{e^{{r_0}-r}}{r^2 (\sinh (r)-1)}-\frac{\alpha }{r+\sinh (r)+1}\right)
 \end{eqnarray}

 Using Equations \eqref{pt2}-\eqref{density2}, we get the following terms:
 \begin{eqnarray}
 \rho+p_r&=&\frac{e^{-r} \left(\alpha  e^r r^2-\sinh (r) \left(\alpha  e^r r^2+e^{{r_0}}\right)-(r+1) e^{{r_0}}\right)}{8 \pi  r (r+\sinh (r)+1)}
 \end{eqnarray}

 \begin{eqnarray}
\rho+p_t&=& -\frac{e^{-r} (r-2) \left(\alpha  \left(-e^r\right) r^2+\sinh (r) \left(\alpha  e^r r^2+e^{{r_0}}\right)+(r+1) e^{{r_0}}\right)}{16 \pi  r^2 (r+\sinh (r)+1)}
 \end{eqnarray}

  \begin{eqnarray}
 \rho+p_r+2p_t&=&-\frac{1}{16 \pi  r^2 (r+\sinh (r)+1)}e^{-r} (r-2) \left(\alpha  \left(-e^r\right) r^2+\sinh (r) \left(\alpha  e^r r^2+e^{{r_0}}\right)\right.\nonumber\\
 &+&\left.(r+1) e^{{r_0}}\right)
 \end{eqnarray}

 \begin{eqnarray}
 \rho-|p_r|&=&\frac{((r-2) \sinh (r)-2 r)}{16 \pi } \left(-\frac{e^{{r_0}-r}}{r^2 (\sinh (r)-1)}-\frac{\alpha }{r+\sinh (r)+1}\right)-\Big| \frac{(r+2) \sinh (r)}{16 \pi }\nonumber\\
 &\times& \left(-\frac{e^{{r_0}-r}}{r^2 (\sinh (r)-1)}-\frac{\alpha }{r+\sinh (r)+1}\right) \Big|
 \end{eqnarray}

\begin{eqnarray}
\rho-|p_t|&=& \frac{((r-2) \sinh (r)-2 r)}{16 \pi } \left(-\frac{e^{{r_0}-r}}{r^2 (\sinh (r)-1)}-\frac{\alpha }{r+\sinh (r)+1}\right)-\Bigg|\frac{1}{16 \pi }\nonumber\\
&\times&\Bigg[(r+2) \left(-\frac{e^{{r_0}-r}}{r^2 (\sinh (r)-1)}-\frac{\alpha }{r+\sinh (r)+1}\right)\Bigg]\Bigg|
\end{eqnarray}

\begin{eqnarray}
p_t-p_r &=&\frac{1}{16 \pi  r^2 (r+\sinh (r)+1)}e^{-r} (r+2) \left(\alpha  \left(-e^r\right) r^2+\sinh (r) \left(\alpha  e^r r^2+e^{r_0}\right)\right.\nonumber\\
&+&\left.(r+1) e^{r_0}\right)
\end{eqnarray}

\begin{eqnarray}
-\rho+p_r+2p_t&=&\frac{e^{-r} \left(\alpha  e^r r^2-\sinh (r) \left(\alpha  e^r r^2+e^{r_0}\right)-(r+1) e^{r_0}\right)}{4 \pi  r^2 (\sinh (r)-1)}
\end{eqnarray}

\begin{figure}
	\centering
	\subfigure[Density ($\rho$) versus $r$ with $\alpha=-1$]{\includegraphics[scale=.65]{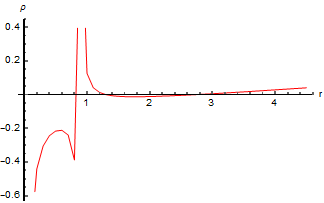}}\hspace{.05cm}
	\subfigure[NEC term $\rho + p_r$ versus $r$ with $\alpha=-1$]{\includegraphics[scale=.65]{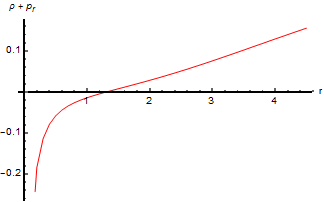}}\hspace{.05cm}
	\subfigure[NEC term $\rho + p_t$ versus $r$ with $\alpha=-1$]{\includegraphics[scale=.65]{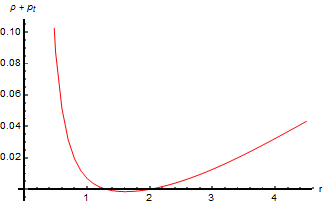}}\hspace{.05cm}
	\subfigure[SEC term $\rho + p_r + 2p_t$ versus $r$ with $\alpha=-1$]{\includegraphics[scale=.65]{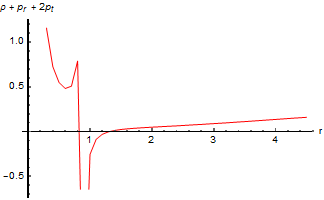}}\hspace{.05cm}
	\subfigure[DEC term $\rho - |p_r|$ versus $r$ with $\alpha=-1$]{\includegraphics[scale=.65]{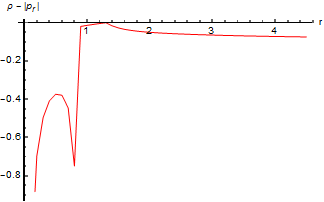}}\hspace{.05cm}
	\subfigure[DEC term $\rho - |p_t|$ versus $r$ with $\alpha=-1$]{\includegraphics[scale=.65]{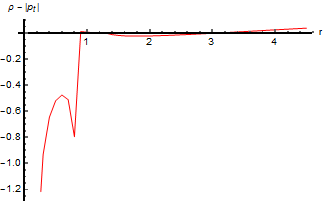}}\hspace{.05cm}
	\subfigure[Anisotropy parameter $\triangle$ versus $r$ with $\alpha=-1$]{\includegraphics[scale=.65]{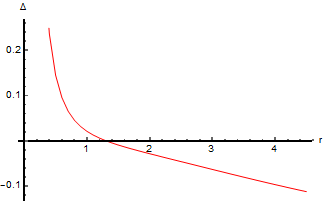}}\hspace{.05cm}
	\subfigure[EoS parameter $\omega$ versus $r$ with $\alpha=-1$]{\includegraphics[scale=.65]{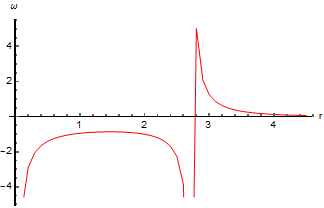}}\hspace{.05cm}
\end{figure}

	\begin{figure}
	\centering
		\subfigure[Stress energy tensor $T$ versus $r$ with $\alpha=-1$]{\includegraphics[scale=.65]{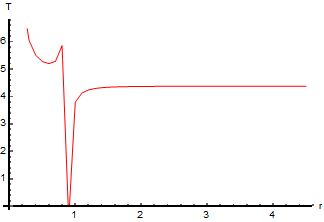}}\hspace{.05cm}
\end{figure}

\section{Energy Conditions}
The important energy conditions are the Null Energy Condition (NEC), Weak Energy Condition (WEC), Strong Energy Condition (SEC) and Dominant Energy Condition (DEC). For any null vector, the null energy condition (NEC) is defined as
$NEC\Leftrightarrow T_{\mu\nu}k^{\mu}k^{\nu}\geq 0$.
Alternately, in terms of the principal pressures NEC is defined as $NEC\Leftrightarrow ~~ \forall i, ~\rho+p_{i}\ge 0$.
For a time-like vector, the weak energy condition (WEC) is defined as $WEC\Leftrightarrow T_{\mu\nu}V^{\mu}V^{\nu}\ge 0$. In terms of the principal pressures, it is defined as $WEC\Leftrightarrow \rho\ge 0;$ and $\forall i,  ~~ \rho+p_{i}\ge 0$. For a time-like vector, the strong energy condition (SEC) is defined as $SEC\Leftrightarrow (T_{\mu\nu}-\frac{T}{2}g_{\mu\nu})V^{\mu}V^{\nu}\ge 0$, where $T$ is the trace of the stress-energy tensor. In terms of the principal pressures, SEC is defined as
$T=-\rho+\sum_j{p_j}$ and $SEC\Leftrightarrow \forall j, ~ \rho+p_j\geq 0, ~ \rho+\sum_j{p_j}\geq 0$.  For any time-like vector, the dominant energy condition (DEC) is defined as $DEC\Leftrightarrow T_{\mu\nu}V^{\mu}V^{\nu}\ge 0$, and
$T_{\mu\nu}V^{\mu}$ is not space-like. In terms of the principal pressures $DEC\Leftrightarrow \rho\ge 0;$ and $\forall i, ~ p_i\in [-\rho, ~+\rho]$.

 In this paper, these conditions are investigated in terms of principal pressures which are as follows:
\begin{itemize}
	\item [(I)] $\rho + p_r\geq 0$, $\rho + p_t\geq 0$ (NEC)
	\item [(II)] $\rho \geq 0$, $\rho + p_r\geq 0$, $\rho + p_t\geq 0$ (WEC)
	\item [(III)] $\rho + p_r\geq 0$, $\rho + p_t\geq 0$,  $\rho + p_r +2p_t\geq 0$ (SEC)
	\item [(IV)] $\rho\geq 0$, $\rho - \lvert p_r\rvert \geq 0$, $\rho - \lvert p_t\rvert \geq 0$ (DEC)
\end{itemize}

A normal matter always satisfies these energy conditions because it possesses positive pressure and positive energy density. The wormholes are non-vacuum solutions of Einstein's field equations and according to Einstein's field theory, they are filled with a matter which is different from the normal matter and is known as exotic matter. This matter does not validate the energy conditions.

\section{Results}
In the present paper, the existence of wormhole solutions is examined in the background of $f(R,T)$ theory of gravity with a novel $f(R,T)$ function defined by $f(R,T)=R+2\alpha \ln(T)$, where $\alpha$ is a constant and $T=-\rho+p_r+2p_t>0$. In Section 3, the field equations are derived and in Section 4, these equations are solved for the following two shape functions defined in the literature:
I. $ b(r)=\frac{r_0\log(r+1)}{\log(r_0+1))}$ and     II. $b(r)=\frac{r}{\exp(r-r_0)}$. In Section 4, the energy condition terms, equation of state parameter, anisotropy parameter and stress energy tensor are also computed in the context of these two shape functions. In this section, the results are discussed and analysed in detail for each type of shape function.
\\

\noindent
\textbf{Case I:} $ b(r)=\frac{r_0\log(r+1)}{\log(r_0+1))}$\\
This case is dealt with in two subcases I(a): $\alpha\geq 0$ and I(b): $\alpha<0$. For each subcase, the results are summarized in Table-1. For positivity of the energy density $\rho$ or any other term, first the positivity of the stress energy tensor $T$ is required. For $\alpha\geq 0$ in Table-1, it can be seen that $T$ is positive only for $r\in(0,0.9)$. Therefore, the model $f(R, T)=R+\alpha\ln (T)$ is physically realistic only near the throat with respect to the shape function $ b(r)=\frac{r_0\log(r+1)}{\log(r_0+1))}$, elsewhere, i.e. outside the throat of the wormhole, the model is not physically realistic.  The energy density is found to be positive for $r\in(0.8,\infty)$. However,  the first NEC term $\rho+p_r$ is negative throughout, so we could say, all the energy conditions are violated for this model. Therefore, we must say that the exotic matter can not be avoided in wormholes modeling in this $f(R, T)=R+\alpha\ln (T)$ gravity, where $\alpha\ge 0$, with respect to the shape function $ b(r)=\frac{r_0\log(r+1)}{\log(r_0+1))}$. Now, for $\alpha<0$, $T$ is positive for $r\in(1,\infty)$, so we could say that the model $f(R, T)=R+\alpha\ln (T)$ is physically realistic for $\alpha<0$ to sustain the wormhole solutions, provided the size of the throat of the wormhole will be more than one, i.e. $r_0>1$. Subsequently, we observed that the energy density $\rho$ as well as $\rho+p_t$ are negative in this range of $r$, so we can say that all energy conditions are violated throughout. This means not only throat, but also the entire geometry of the wormhole is filled with exotic matter. Hence, we conclude that the exotic matter can not be avoided for the construction of wormhole geometry with this particular choice of $f(R, T)=R+\alpha\ln (T)$ and $ b(r)=\frac{r_0\log(r+1)}{\log(r_0+1))}$.
\\

\noindent
\textbf{Case II:} $b(r)=\frac{r}{\exp(r-r_0)}$\\
This case also involves two subcases with respect to the parameter $\alpha$ and for each subcase the results are summarized in Table-2. For $\alpha\geq 0$, $T>0$ for $r\in(0,0.8)$.  Therefore, the model $f(R, T)=R+\alpha\ln (T)$ is physically realistic only for a very small range of $r$ with respect to the shape function $ b(r)=\frac{r}{\exp(r-r_0)}$ and elsewhere the model is not physically realistic. For the same range of $r$, $\rho<0$ and $\rho+p_r<0$. Thus, our second choice of shape function could not avoid the presence of exotic matter for $\alpha\geq 0$ near the throat of the wormhole as well. Now, for $\alpha<0$, $T>0$ for $r\in (0,\infty)$. Hence, the model $f(R, T)=R+\alpha\ln (T)$ is physically realistic throughout the geometry of the wormhole.
The first NEC term $\rho+p_r>0$ for $r\in (1.3,\infty)$ and the second NEC term $\rho+p_t>0$ for $r\in (0,1.2]\cup[2,\infty)$. Thus, NEC is satisfied for $r\in[2,\infty)$. This estimate indicates, we could avoid the presence of exotic matter for the construction of wormholes, provided the size of the throat of the wormhole must be greater than two ($r_0>2$). Further, the energy density is observed to be positive for  $r\in[0.8,1.3]\cup(2.7,\infty)$. This implies the validation of WEC for  $r\in (2.7,\infty)$. Now, SEC term $\rho+p_r+2p_t>0$ for $r\in(0,0.8]\cup[1.3,\infty)$. For the satisfaction of SEC, the terms $\rho+p_r$, $\rho+p_t$ and $\rho+p_r+2p_t$ should be positive and these three terms are positive for $r\in[2,\infty)$. Hence, SEC is valid for $r\in[2,\infty)$. Furthermore, the first DEC term is negative throughout, hence DEC is violated everywhere. Thus, all NEC, WEC and SEC are satisfied for $r\in (2.7,\infty)$. In this range, where energy conditions are fulfilled, the  anisotropy parameter is negative and the equation of state  parameter is positive. This indicates the presence of attractive geometry filled with ordinary matter. In particular, for $\alpha = -1$, the results are plotted in Figs. 1(a) to 1(i). Thus, it also shows that the presence of exotic matter can be avoided completely if the radius of the throat is chosen greater than $2.7$.   So, this subcase for $\alpha<0$ with $b(r)=\frac{r}{\exp(r-r_0)}$ is favourable and provides the wormhole solutions with desired properties.

\begin{table}[!h]
	\centering
	\caption{Summary of results for $ b(r)=\frac{r_0\log(r+1)}{\log(r_0+1))}$}
	\begin{tabular}{|c|c|l|l|}
		\hline
		S.No.& Terms& $\alpha\geq 0$& $\alpha<0$\\
		\hline
		1 & $\rho$ & $>0$, for $r\in(0.8,\infty)$&$>0$, for $r\in(0.8,1)$\\
		&       & $<0$, for $r\in(0,0.8]$&$<0$, for $r\in(0,0.8]\cup[1,\infty)$\\
				\cline{1-4}
		2 & $\rho+p_r$ &  $<0$, for $r\in(0,\infty)$& $>0$, for $r\in[1,\infty)$\\
		&       &   &  $<0$, for $r\in(0,1)$\\
			\cline{1-4}
		3 & $\rho+p_t$ & $>0$, for $r\in (0,\infty)$&$>0$, for $r\in(0,1]$\\
		&            & &$<0$, for $r\in (1,\infty)$\\
				\cline{1-4}
		4 &  $\rho+p_r+2p_t$ & $>0$, for $r\in(0, 0.8]$ & $>0$, for $r\in(0, 0.8]\cup (1,\infty)$ \\
					&            & $<0$, $r\in(0.8,\infty)$&$<0$, for $r\in (0.8,1]$\\
		\cline{1-4}
		5 &  $\rho-|p_r|$ & $<0$, for $r\in(0, \infty)$& $<0$, for $r\in(0, \infty)$\\
		\cline{1-4}
		6 &  $\rho-|p_t|$ & $>0$, for $r\in(0.8, \infty)$ & $>0$, for $r\in(0.9,1]$\\
		&               & $<0$, for $r\in (0,0.8]$& $<0$, for $r\in(0,0.9]\cup(1,\infty)$\\
					\cline{1-4}
		7 &  $\triangle$ & $>0$, for $r\in(0, \infty)$ & $>0$, for $r\in(0, 1]$\\
		&&		&	$<0$, for $r\in(1,\infty)$\\
		\cline{1-4}
		8 &  $\omega$& $<0$, for $r\in(0,\infty)$& $<0$, for $r\in(0,\infty)$\\
	\hline
	9 &  $T$ & $>0$, for $r\in(0,0.9)$& $>0$, for $r\in(1,\infty)$\\
	&               & $<0$, for $r\in [0.9,\infty)$& $<0$, for $r\in(0,1]$\\
	\hline
			\end{tabular}
\end{table}

\begin{table}[!h]
	\centering
	\caption{Summary of results for $b(r)=\frac{r}{\exp(r-r_0)}$}
	\begin{tabular}{|c|c|l|l|}
		\hline
		S.No.& Terms& $\alpha\geq 0$& $\alpha<0$\\
		\hline
		1 & $\rho$ & $>0$, for $r\in(0.8,2.8)$&$>0$, for $r\in[0.8,1.3]\cup(2.7,\infty)$\\
		&       & $<0$, for $r\in(0,0.8]\cup[2.8,\infty)$&$<0$, for $r\in(0,0.8)\cup(1.3,2.7]$\\
				\cline{1-4}
		2 & $\rho+p_r$ &  $<0$, for $r\in(0,\infty)$& $>0$, for $r\in(1.3,\infty)$\\
		&       &   &  $<0$, for $r\in(0,1.3]$\\
			\cline{1-4}
		3 & $\rho+p_t$ & $>0$, for $r\in (0,2)$&$>0$, for $r\in(0,1.2]\cup[2,\infty)$\\
		&            & $<0$, $r\in[2,\infty)$&$<0$, for $r\in (1.2,2)$\\
				\cline{1-4}
		4 &  $\rho+p_r+2p_t$ & $>0$, for $r\in(0, 0.9)$ & $>0$, for $r\in(0, 0.8]\cup [1.3,\infty)$ \\
					&            & $<0$, $r\in[0.9,\infty)$&$<0$, for $r\in (0.8,1.3)$\\
		\cline{1-4}
		5 &  $\rho-|p_r|$ & $<0$, for $r\in(0, \infty)$& $<0$, for $r\in(0, \infty)$\\
		\cline{1-4}
		6 &  $\rho-|p_t|$ & $>0$, for $r\in(0.8, 2)$ & $>0$, for $r\in[0.9,1.3]\cup(3,\infty)$\\
		&               & $<0$, for $r\in [0,0.8]\cup[2,\infty)$& $<0$, for $r\in(0,0.9)\cup(1.3,3]$\\
					\cline{1-4}
		7 &  $\triangle$ & $>0$, for $r\in(0, \infty)$ & $>0$, for $r\in(0, 1.3]$\\
		&&		&	$<0$, for $r\in(1.3,\infty)$\\
		\cline{1-4}
		8 &  $\omega$& $>0$, for $r\in(2.7,\infty)$& $>0$, for $r\in(2.7,\infty)$\\
		&&  $<0$, for $r\in(0,2.7]$& $<0$, for $r\in(0,2.7]$\\
						\hline
						9 &  $T$ & $>0$, for $r\in(0,0.8)$& $>0$, for $r\in(0,\infty)$\\
	&               & $<0$, for $r\in [0.8,\infty)$& \\
	\hline
			\end{tabular}
\end{table}

\begin{table}[!h]
	\centering
	\caption{Summary for the satisfaction/ violation of energy conditions}
	\begin{tabular}{|c|c|l|l|}
		\hline
		S.No.& Energy condition& $ b(r)=\frac{r_0\log(r+1)}{\log(r_0+1))}$& $b(r)=\frac{r}{\exp(r-r_0)}$\\
		\hline\hline
		1 & NEC & Violated $\forall$ $r$, $\alpha$&Satisfied for $r\in[2,\infty)$, $\alpha<0$\\
			\hline
		2 & WEC & Violated $\forall$ $r$, $\alpha$&Satisfied for $r\in (2.7,\infty)$, $\alpha<0$\\
			\hline
		3 & SEC & Violated $\forall$ $r$, $\alpha$&Satisfied for $r\in [2,\infty)$, $\alpha<0$\\
		\hline
		4 & DEC & Violated $\forall$ $r$, $\alpha$&Violated $\forall$ $r$, $\alpha$\\
		\hline
        5 & WEC, SEC & Violated $\forall$ $r$, $\alpha$&Satisfied for $r\in (2.7,\infty)$, $\alpha<0$\\
		
			\hline
			
			\end{tabular}
\end{table}

 \section{Conclusion}
In general relativity, traversable wormholes exist in the presence of exotic matter that does not satisfy the null energy condition. In the present work, the generalized theory of $f(R,T)$ gravity is taken into account to explore the existence of wormhole solutions. This modified theory provides the material correction by including a factor in terms of stress energy tensor. As a result, it may lead to the satisfaction of null energy condition. To explore this possibility, at first, a new form of $f(R,T)$ function  is defined as $f(R,T)=R + 2\alpha \ln(T)$, where $\alpha$ is a constant and $T=-\rho+p_r+2p_t>0$ and the field equations are derived. To solve the field equations, the equation of state is taken as $p_t=\omega \rho$ and the pressures are considered to be related as $p_r=\sinh (r) p_t$. The derived field equations are non-linear and they demand the choice of shape function. In this work, we have chosen the two shape functions introduced in the literature: I. $ b(r)=\frac{r_0\ln(r+1)}{\ln(r_0+1))}$ and II. $b(r)=\frac{r}{\exp(r-r_0)}$. Subsequently, we have analysed the presence of exotic matter,  energy conditions and geometric configuration for these shape functions and compared the results. It is found that shape function $b(r)=\frac{r_0\ln(r+1)}{\ln(r_0+1))}$ is not a suitable choice to avoid the presence of exotic matter near the throat of the wormhole. However, the shape function $b(r)=\frac{r}{\exp(r-r_0)}$ with $\alpha<0$ provides the wormhole solutions without the support of exotic matter. These solutions obey NEC and also WEC and SEC for $r>2.7$. These are filled with ordinary matter having attractive geometry. It is concluded that the energy conditions are violated throughout or the presence of exotic matter can not be avoided, in case of general relativity, i.e. for $\alpha=0$. However, in the framework of $f(R,T)=R + 2\alpha \ln(T)$ gravity with $\alpha<0$, the wormholes with the radius of the throat $>2.7$ do not contain any type of exotic matter and they avoid the violation of energy conditions completely for the choice of shape function $b(r)=\frac{r}{\exp(r-r_0)}$.

\section*{Acknowledgement}
The authors are very much thankful to the reviewers and editors for their constructive comment to improve the work significantly. 



\begin{thebibliography}{1}

\bibitem{Wheeler1}J. A. Wheeler, Geometrodynamics (Academic, New York, 1962).

\bibitem{Misner1973} C. W. Misner, K. S. Thorne and J. A. Wheeler, Gravitation, Freeman, San Francisco, 1973.

\bibitem{Morris1} M. S. Morris and K. S. Thorne, Am. J. Phys. \textbf{56} (1988) 395.

\bibitem{eins-ros}A. Einstein and N. Rosen, Ann. Phys. \textbf{2} (1935) 242.

\bibitem{Wheeler}J. A. Wheeler, Geometrodynamics (Academic, New York,
1962).

\bibitem{Hawking}S. W. Hawking, Phys. Rev. D \textbf{37}  (1988) 904.

\bibitem{Morris2}M. S. Morris, K. S. Thorne and U. Yurtsever, Phys. Rev. Lett. \textbf{61} (1988) 1446.

\bibitem{Frolov} V. P. Frolov and I. D. Novikov, Phys. Rev. D \textbf{42} (1990)  1057.

\bibitem{Mehdizadeh04} M. R. Mehdizadeh, M. Kord Zangeneh and F. S. N. Lobo, Phys. Rev. D \textbf{91} (2015) 084004.

\bibitem{Zangeneh49} M. K. Zangeneh, F. S. N. Lobo and M. H. Dehghani, Phys. Rev. D \textbf{92} (2015) 124049.

\bibitem{Mehdizadeh14}M. R. Mehdizadeh and F. S. N. Lobo, Phys. Rev. D \textbf{93} (2016) 124014.

\bibitem{Shaikh11}R. Shaikh and S. Kar, Phys. Rev. D \textbf{94} (2016) 024011.

\bibitem{Moradpour66}H. Moradpour, N. Sadeghnezhad and S. H. Hendi, Can. J. Phys, \textbf{95} (2017) 1257.



\bibitem{Tsujikawa} S. Tsujikawa, Lect. Notes Phys. \textbf{800} (2010) 99.

\bibitem{Capozziello}S. Capozziello and M. de Laurentis, Phys. Rep. \textbf{509} (2011) 167.

\bibitem{Nojiri}S. Nojiri and S. D. Odintsov, Phys. Rept. \textbf{505} (2011) 59.

\bibitem{Clifton}T. Clifton, P. G. Ferreira, A. Padilla and C. Skordis, Physics Reports \textbf{513} (2012) 1.

\bibitem{Berti}E. Berti, et al., Class. Quant. Grav. \textbf{32} (2015) 243001.

\bibitem{Buchdahl}H. A. Buchdahl, Mon. Not. R. Astron. Soc. \textbf{150} (1970) 1.

\bibitem{Star} A. A. Starobinsky, Phys. Lett. B \textbf{91} (1980) 99.

\bibitem{Bertolami}O. Bertolami, C. G. Böhmer, T. Harko, and F. S. N. Lobo, Phys. Rev. D \textbf{75} (2007) 104016.

\bibitem{Nojiris}S. Nojiri and S. D. Odintsov, Int. J. Geom. Methods Mod. Phys. \textbf{4} (2007) 115.

\bibitem{Bertolami1}O. Bertolami, P. Fraz$\tilde{a}$o and J. P\'{a}ramos, Phys. Rev. D \textbf{81} (2010) 104046.




\bibitem{Bamba} K. Bamba, S. Capozziello, S. Nojiri and S. D. Odintsov, Astrophys. Space Sci. \textbf{342} (2012) 155.


\bibitem{Nojiri001} S. Nojiri, S. D. Odintsov and M. Sami,
Phys. Rev. D \textbf{74} (2006) 046004.

\bibitem{Shirasaki} Y. Shirasaki, Y. Komiya, M. Ohishi and Y. Mizumoto,
Publ. Astron. Soc. Jap. \textbf{68} (2016) 23.
\bibitem{Capozziello001} S. Capozziello, T. Harko, T. S. Koivisto, F. S. N. Lobo and G. J. Olmo,
J. Cosm. Astrop. Phys. \textbf{2013} (2013) 024.
\bibitem{Rodrigues01} D. C. Rodrigues, P. L. de Oliveira, J. C. Fabris and G. Gentile,
Month. Not. Roy. Astron. Soc. \textbf{445} (2014) 3823.






\bibitem{Capozziello91} 	
S. Capozziello, C. A. Mantica and L. G. Molinari, Int. J. Geom. Meth. Mod. Phys. \textbf{16} (2018) 1950008.



\bibitem{Bombacigno21} 	
F. Bombacigno and G. Montani, Eur. Phys. J. C \textbf{79} (2019) 405.



\bibitem{Sbis} 	
F. Sbis\`{a}, O. F. Piattella and S. E. Jor\'{a}s, Phys. Rev. D \textbf{99} (2019) 104046.

\bibitem{Chen} 	
L. Chen, Phys. Rev. D \textbf{99} (2019) 064025.

\bibitem{Elizalde26} 	
E. Elizalde, S. D. Odintsov, V. K. Oikonomou and T. Paul, JCAP \textbf{1902} (2019) 017.

\bibitem{Elizalde06} 	
E. Elizalde, S. D. Odintsov, T. Paul and D. S\'{a}ez-Chill\'{o}n G\'{o}mez, Phys. Rev. D \textbf{99} (2019) 063506.

\bibitem{Astashenok1} 	
A. V. Astashenok, K. Mosani, S. D. Odintsov and G. C. Samanta, Int. J. Geom. Meth. Mod. Phys. \textbf{16} (2019)  1950035.

\bibitem{Miranda} 	
T. Miranda, C. Escamilla-Rivera, O. F. Piattella and J. C. Fabris, JCAP \textbf{2019} (2019) 028.

\bibitem{Nascimento} 	
J. R. Nascimento, G. J. Olmo, P. J. Porfirio, A. Yu. Petrov and A. R. Soares, Phys. Rev. D \textbf{99} (2019) 064053.

\bibitem{Odintsov20} 	
S. D. Odintsov and V. K. Oikonomou, Phys. Rev. D \textbf{99} (2019) 064049.

\bibitem{Odintsov1} 	
S. D. Odintsov and V. K. Oikonomou, Class. Quant. Grav. \textbf{36} (2019) 065008.

\bibitem{Nojiri56} 	
S. Nojiri, S. D. Odintsov and V. K. Oikonomou, Nucl. Phys. B \textbf{941} (2019) 11.

\bibitem{Parth} 	
P. Shah and G. C. Samanta, Eur. Phys. J. C \textbf{79} (2019) 414.


\bibitem{gninjp} G. C. Samanta and N. Godani, Ind. J. Phys. (2019) Doi:10.1007/s12648-019-01565-w.





\bibitem{Chiba} T. Chiba,
Phys. Lett. B \textbf{575}, 1 (2003).
\bibitem{Nojiri21} S. Nojiri and S. D. Odintsov,
Phys. Lett. B \textbf{659}, 821 (2008).

\bibitem{harko}T. Harko, F. S. N. Lobo, S. Nojiri, and S. D. Odintsov., Phys. Rev. D \textbf{84} (2011)  024020.


\bibitem{Houndjo1} 	
M. J. S. Houndjo, Int. J. Mod. Phys. D \textbf{21} (2012) 1250003.

\bibitem{Sharif01}M. Sharif and M. Zubair, JCAP \textbf{1203} (2012) 028.







\bibitem{Jamil1} 	
M. Jamil, D. Momeni,  M. Raza and R. Myrzakulov, Eur. Phys. J. C \textbf{72} (2012) 1999.

\bibitem{Alvarenga01} 	
F. G. Alvarenga, A. de la Cruz-Dombriz, M. J. S. Houndjo, M. E. Rodrigues and D. S\`{a}ez-G\'{o}mez, Phys. Rev. D \textbf{87} (2013) 103526.

\bibitem{Santos1} 	
A. F. Santos, Mod. Phys. Lett. A \textbf{28} (2013) 1350141.


\bibitem{Samanta} 	
G. C. Samanta, Int. J. Theor. Phys. \textbf{52} (2013) 2303.

\bibitem{Shabani} 	
H. Shabani and M. Farhoudi, Phys. Rev. D \textbf{88} (2013) 044048.

\bibitem{Naidu}  	
R. L. Naidu, D. R. K. Reddy, T. Ramprasad and K. V. Ramana, Astrophys. Space Sci. \textbf{348} (2013) 247.

\bibitem{Samanta3}
G. C. Samanta and S. N. Dhal, Int. J. Theor. Phys. \textbf{52} (2013) 1334.

\bibitem{Chandel} 	
S. Chandel and S. Ram, Ind. J. Phys. \textbf{87} (2013) 1283.

\bibitem{Samanta1} 	
G. C. Samanta, Int. J. Theor. Phys. \textbf{52} (2013) 2647.

\bibitem{Shabani001} 	
H. Shabani and M. Farhoudi, Phys. Rev. D \textbf{90} (2014) 044031.


\bibitem{Samanta2} 	
G. C. Samanta, S. Jaiswal and S. K. Biswal, Eur. Phys. J. Plus \textbf{129} (2014) 48.

\bibitem{Moraes01} 	
P. H. R. S. Moraes, Eur. Phys. J. C \textbf{75} (2015) 168.

\bibitem{Noureen} 	
I. Noureen and M. Zubair, Eur. Phys. J. C \textbf{75} (2015) 62.


\bibitem{Farasat}
M. Farasat Shamir, Eur. Phys. J. C \textbf{75} (2015) 354.

\bibitem{Mirza} 	
B. Mirza and F. Oboudiat, Int. J. Geom. Meth. Mod. Phys. \textbf{13} (2016) 1650108.


\bibitem{Correa1}
R. A. C. Correa and P. H. R. S. Moraes, Eur. Phys. J. C \textbf{76} (2016) 100.

\bibitem{Ramesh} 	
G. Ramesh and S. Umadevi, Astrophys. Space Sci. \textbf{361} (2016) 2.

\bibitem{Moraes0001} 	
P. H. R. S. Moraes and R. A. C. Correa, Astrophys. Space Sci. \textbf{361} (2016) 91.

\bibitem{Moraes9}  	
P. H. R. S. Moraes, Jose D. V. Arba\~{n}il and M. Malheiro, JCAP \textbf{1606} (2016) 005.

\bibitem{Zaregonbadi} 	
R. Zaregonbadi, M. Farhoudi and N. Riazi, Phys. Rev. D \textbf{94} (2016) 084052.

\bibitem{Rahaman1} 	
A. Das, F. Rahaman, B. K. Guha and S. Ray, Eur. Phys. J. C \textbf{76} (2016) 654.

\bibitem{Mishra} 	
B. Mishra, S. Tarai and S. K. Tripathy, Adv. High Energy Phys. \textbf{2016} (2016) 8543560.

\bibitem{Yousaf} 	
Z. Yousaf, K. Bamba and M. Z. ul Haq Bhatti, Phys. Rev. D \textbf{93} (2016) 124048.

\bibitem{Samanta5} 	
G. C. Samanta, R. Myrzakulov and Parth Shah, Z. Naturforsch. A \textbf{72} (2017) 365.





\bibitem{Aditya} 	
Y. Aditya and D. R. K. Reddy, Astrophys. Space Sci. \textbf{364} (2019) 3.

\bibitem{Elizalde} 	
E. Elizalde and M. Khurshudyan, Phys. Rev. D \textbf{99} (2019) 024051.

\bibitem{Ordines} 	
T. M. Ordines and E. D. Carlson, Phys. Rev. D \textbf{99} (2019) 104052.

\bibitem{Pavlovic} 	
P. Pavlovic and M. Sossich, Eur. Phys. J. C \textbf{75} (2015) 117.

\bibitem{Habib92}	
S. Habib Mazharimousavi and M. Halilsoy, Mod. Phys. Lett. A \textbf{31} (2016) 1650192.



\bibitem{Lobo1} F. S. N. Lobo and M. A. Oliveira, Phys. Rev. D \textbf{80} (2009) 104012.

\bibitem{Saiedi} H. Saiedi and B. N. Esfahani, Mod. Phys. Lett. A \textbf{26} (2011) 1211.

\bibitem{Eiroa1} 	
E. F. Eiroa and G. F. Aguirre, Eur. Phys. J. C \textbf{76} (2016) 132.


\bibitem{peter}P. K. F. Kuhfittig, Ind. J. Phys. \textbf{92} (2018) 1207.

\bibitem{ns} N. Godani and G. C. Samanta, Int. J. Mod. Phys. D \textbf{28}  (2018) 1950039.

\bibitem{godani1} G. C. Samanta, N. Godani and K. Bamba, arXiv:1811.06834v1  [gr-qc] (2018).

\bibitem{ngmpla} N. Godani and G. C. Samanta, Mod. Phys. Lett. A (2019)     DOI: 10.1142/S0217732319502262.

\bibitem{gnmpla} G. C. Samanta and N. Godani,  Mod. Phys. Lett. A (2019)  DOI:	10.1142/S0217732319502249.

\bibitem{gnepjc} G. C. Samanta and N. Godani, Eur. Phys. J. C \textbf{79} (2019) 623.

\bibitem{Azizi}
T. Azizi, Int. J. Theo. Phys. \textbf{52} (2013) 3486.

\bibitem{Zubair} M. Zubair, S. Waheed  and Y. Ahmad, Eur. Phys. J. C \textbf{76} (2016) 444.

\bibitem{Zubair80} 	
M. Zubair, G. Mustafa, S. Waheed and G. Abbas, Eur. Phys. J. C \textbf{77} (2017) 680.

\bibitem{Yousaf63} 	
Z. Yousaf, M. Ilyas and M. Z. Bhatti, Mod. Phys. Lett. A \textbf{32} (2017) 1750163.


\bibitem{Moraes2} P. H. R. S. Moraes, R. A. C. Correa and R. V. Lobato, J. Cosmol. Astropart. Phys. \textbf{2017} (2017).

\bibitem{Elizalde25}  	
E. Elizalde and M. Khurshudyan, Phys. Rev. D \textbf{98} (2018) 123525.

\bibitem{Bhatti}M. Z. Bhatti, Z. Yousaf and M. Ilyas, J. Astrophys. Astr. \textbf{39} (2018) 69.

\bibitem{Elizalde51} 	
E. Elizalde and M. Khurshudyan, Phys. Rev. D \textbf{99} (2019) 024051.

\bibitem{Moraes1} P. H. R. S. Moraes, W. de Paula1 and R. A. C. Correa, Int. J. Mod. Phys. D \textbf{28} (2019) 1950098.

\bibitem{Sharif} M. Sharif and I. Nawazish, Ann.  Phys. \textbf{400} (2019) 37.

\bibitem{Landau} L. D. Landau and E. M. Lifshitz, The Classical Theory of Fields, Butterworth-Heinemann, Oxford (1998).


\bibitem{Linder} E. V. Linder, Phys. Rev. D 80 (2009) 123528.

\bibitem{Campista}M. Campistaa, B. Santosa, J. Santosb and J. S. Alcaniza, Phys. Lett. B 699 (2011) 320.

\bibitem{Bamba1}K. Bamba, A. Lopez-Revelles, R. Myrzakulov, S. D. Odintsov and L. Sebastiani, Class. Quant. Grav. 30 (2013) 015008.

\bibitem{Odintsov}S. D. Odintsova, V. K. Oikonomouc and L. Sebastiani, Nucl. Phys. B 923 (2017) 608.
\end{thebibliography}
\end{document}